\documentclass[letterpaper, 
    ,final            % use final for the camera ready runs
  ]
  {aipproc}

\usepackage{natbib}

\layoutstyle{6x9}

\newcommand{\LISA}{{\em LISA}}
\newcommand{\SgrA}{Sgr\,A$^\ast$}
\newcommand {\MBH}{\ensuremath{M_{\mathrm{BH}}}}
\newcommand {\Mstar}{\ensuremath{M_{\ast}}}
\newcommand {\Msun}{\ensuremath{M_{\odot}}}

\newcommand {\peryr}{\ensuremath{\mathrm{yr}^{-1}}}

\newcommand{\gw}{gravitational waves}

\newcommand{\rem}[1]{} % pour mettre des commentaires en ligne...

\begin{document}

\title{
       Captures of stars by a massive black hole:
       Investigations in numerical stellar dynamics.}%{
            %Stellar captures by a massive black hole}

\author{M. Freitag}{
  address={Astronomisches Rechen-Institut, M\"onchhofstrasse 12-14, D-69120 Heidelberg, Germany},
  email={freitag@ari.uni-heidelberg.de}
}

%\author{<author2>}{
%  address={<common address for author2 and author3>}
%}

%\author{<author3>}{
%  address={<common address for author2 and author3>}
%  ,altaddress={<author1 address>} % additional visiting address
%}

\begin{abstract}
 
Among the astrophysical systems targeted by {\em LISA}, stars on
relativistic orbits around massive black holes (MBHs) are particularly
promising sources. Unfortunately, the prediction for the number and
characteristics of such sources suffers from many uncertainties.
Stellar dynamical Monte Carlo simulations of the evolution of galactic
nucleus models allow more realistic estimates of these quantities. The
computations presented here strongly suggest that the closest such
extreme mass-ratio binary to be detected by {\em LISA} could be a
low-mass MS star (MSS) orbiting the MBH at the center of our Milky
Way. Only compact stars contribute to the expected detections from
other galaxies because MSSs are disrupted by tidal forces too early.

\end{abstract}

\maketitle

%%%%%%%%%%%%%%%%%%%%%%%%%%%%%%%%%%%%%%%%%%%%
%% MAINMATTER
%%%%%%%%%%%%%%%%%%%%%%%%%%%%%%%%%%%%%%%%%%%%
\section{Introduction}

An object of stellar mass orbiting a MBH with a mass below
$10^7\,\Msun$ would become an ideal source of {\gw} (GW) for {\LISA}
in the last few years before plunge through the horizon, if compact
enough to withstand the tidal forces. Such systems will be detectable
to distances as large as a few hundreds Mpc. From the gravitational
signal emitted by the stars as they spiral in, precise information
about the mass and spin of MBHs can be obtained
\cite{Thorne98}.

While considerable progress has been achieved in computing the orbit of a
small body around and its GW signal \cite{GK02,GHK02}, our
understanding of the other aspects of the problem are still far from
satisfactory. At one end, astronomers will have to detect and measure
the parameters of such sources by using signal--processing algorithms
still to be devised. On the other end of the process, Nature has to
create sources by sending stars onto relativistic orbits, through
processes and with rates still debated \cite{Sigurdsson03}. The usual
conservative approach --also applied here-- is to assume that galactic
nuclei are spherical and to rely on 2-body relaxation to bring stars
onto ``capture orbits'', for which, by definition, GW emission is the
main agent of evolution.

\section{Simulations}

To simulate the relaxational stellar dynamics of spherical galactic
nuclei, we rely on our Monte Carlo (MC) code which, besides captures,
includes all the important physics: cluster self-gravity, 2-body
relaxation, stellar collisions, tidal disruptions and stellar
evolution \cite{FB01a,FB02b}.  The specific advantage of the MC
approach is that it treats the various aspects of the stellar dynamics
(collisions, mass-segregation\ldots) in a self-consistent
way. Furthermore, the simulations do not only provide us with capture
rates but with the distribution of stellar and orbital parameters of
captured objects. Among the downsides are the important statistical
noise due to the rarity of capture events and the need to use a number
of particles much lower than the number of stars.  Also, as the
simulation proceeds, the structure of the cluster evolves, making it
difficult to get ``instantaneous'' rates corresponding to a well
defined nucleus model.

We concentrate here on a model set to represent the nucleus
of the Milky Way \cite{GPEGO00}, with a central BH of mass $2.6\times
10^6\,M_\odot$ and a total mass in stars of $8.67\times
10^7\,M_\odot$. A stellar population with a standard initial mass
function and an age of 10\,Gyrs is assumed. There is no initial 
mass segregation. Various simulations were carried out with $2-6\times
10^6$ particles. We refer to \cite{Freitag03} for more details about
these simulations.

In the MC runs, we detect captures by comparing the time scale for
inspiral through GW emission, $T_\mathrm{GW}$ \cite{Peters64} to the
time required by relaxation to induce a significant change of
pericenter distance, $T_\mathrm{mod}$. If
$T_\mathrm{GW}<T_\mathrm{mod}$, the star is considered captured
because, in most such cases, relaxation will not be able to move it to
a safer orbit (for ideas to implement a more accurate criterion, see
\cite{AH03}).

We bracket the effects of tidal interactions between the MBH and MSSs
by adopting either an optimistic or a pessimistic prescription (from
the point of view of the severity of the decrease in the number of
detectable {\SgrA} GW sources). The (over-)optimistic scenario is to
neglect all tidal interactions until the star enters the Roche zone,
of radius $R_\mathrm{R}=R_\ast(2\MBH/\Mstar)^{1/3}$, at which point
it is completely disrupted in one pericenter passage. In the
(over-)pessimistic approach, one considers the cumulative energy
transfered to the star by tides, $E_\mathrm{tid}$, assuming a
parabolic orbit, an $n=1.5$ polytropic structure
\cite{MMMDT87}, complete conversion into heat and no radiation of this
energy. The star is considered lost (as a source) when
$E_\mathrm{tid}$ reached 20\,\% of its self-binding energy.

\section{Results}

\begin{figure}
  \includegraphics[width=.72\textwidth,height=.43\textheight]{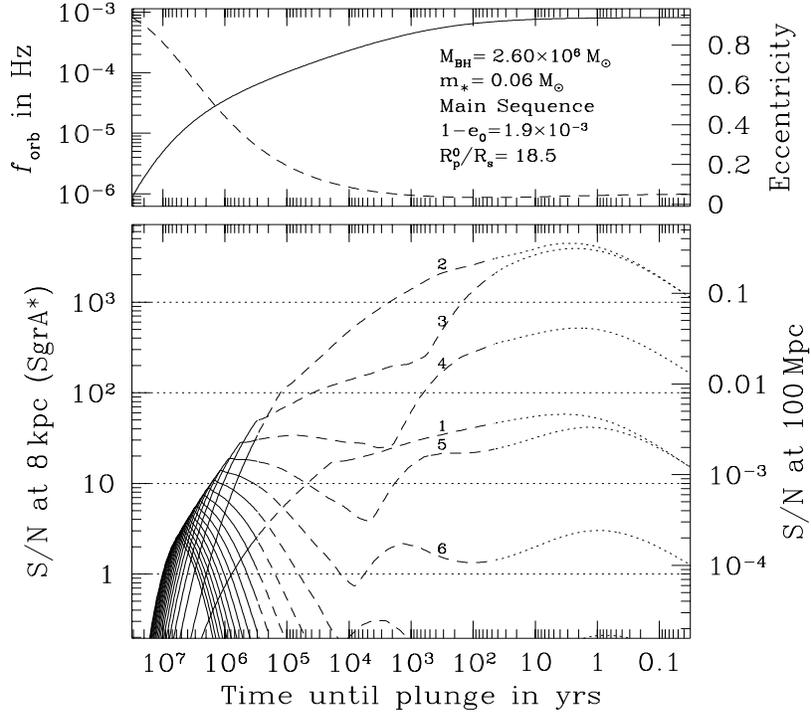}

  \caption{Capture of a low-mass MSS. The top panel shows the evolution
  of the orbital frequency (solid line, left ordinate axis) and
  eccentricity (dashes, right axis), as a function of the time left to
  reach the last stable orbit. We indicate the eccentricity, $e_0$,
  and pericenter distance, $R_\mathrm{p}^0$ (in units of the
  Schwarzschild radius, $R_\mathrm{S} = 2.5\times10^{-7}$\,pc), when
  capture was detected during the MC simulation. On the lower panel,
  we plot the {\LISA} signal-to-noise ratio for the 20 first
  harmonics, of frequencies $n\cdot f_\mathrm{orb}$, of the
  quadrupolar component of gravitational radiation. The curves are
  labelled with their $n$ values. The mission was assumed to last only
  one year. The left ordinate axis corresponds to the Galactic center,
  the right axis to a galaxy at a distance of 100\,Mpc. On the dotted
  segments, the pericentre distance is inside the Roche zone, so the
  MSS is certainly destroyed. On the dashed segments, the star may
  have already suffered from strong tidal damage, according to our
  pessimistic assumption (see text). Even considering only the solid
  segments, which are robust with respect to the effects of tides, one
  sees that this source would have a {\LISA} $\mathrm{S}/\mathrm{N}\ge
  10$ during of order 1--2 million years, if situated at the MW
  center. Note that only considering the 20 first harmonics leads to a
  slight underestimate of the time spent at $\mathrm{S}/\mathrm{N}\ge
  1$. On the other hand, MS sources in distant galaxies clearly never
  reach detectable amplitudes.}

  \label{fig:MSCapt}
\end{figure}

\begin{figure}
  \includegraphics[width=.72\textwidth,height=.43\textheight]{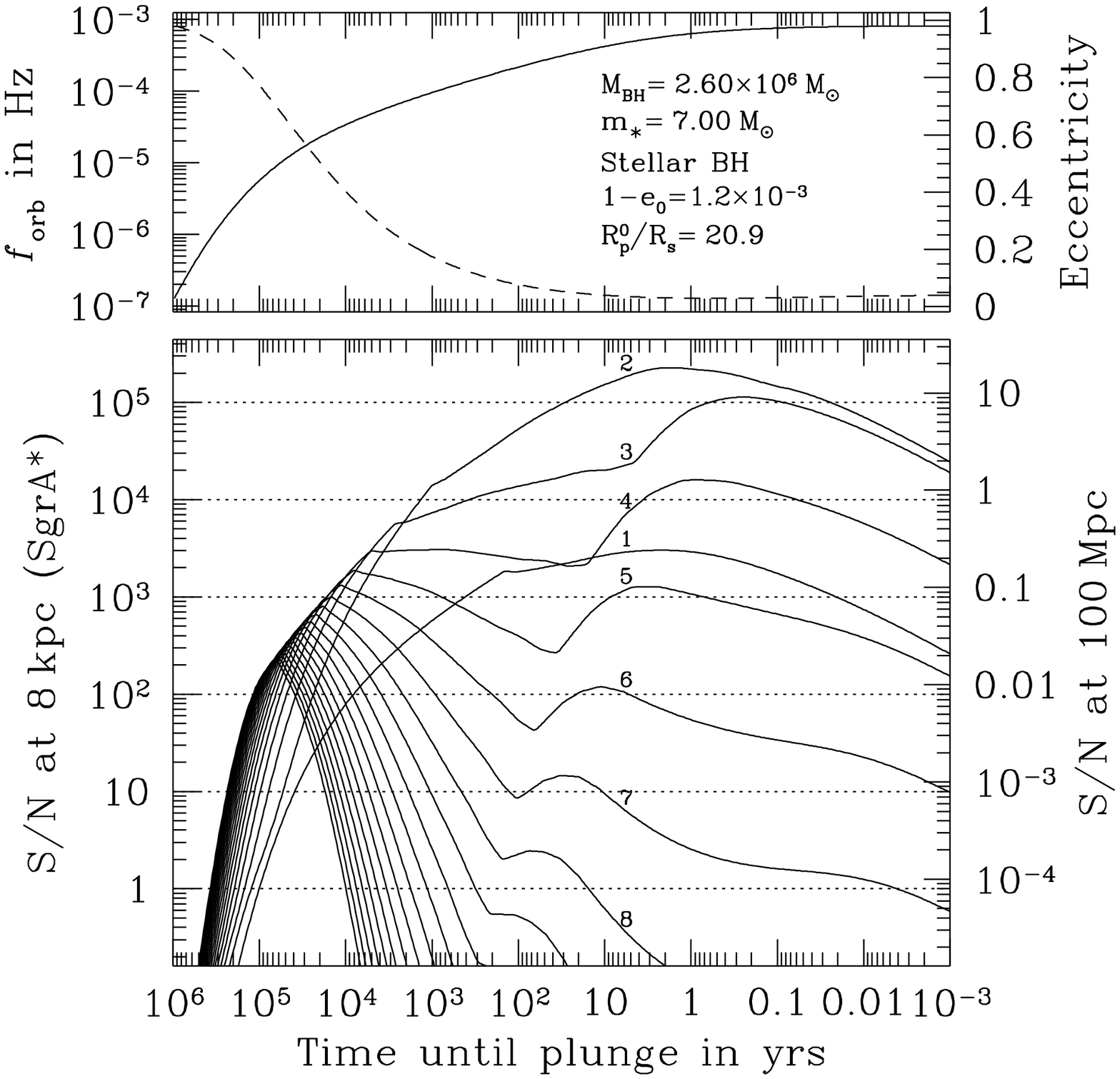}

  \caption{Similar to Figure~\ref{fig:MSCapt} but for a SBH. If
  situated at the Galactic center (\SgrA), the captured object would
  emit with $\mathrm{S}/\mathrm{N}\ge 10$ for a few $10^5$\,yrs (up to
  $10^6$\,yrs if higher harmonics are considered). Given a capture
  rate lower than $10^{-7}\,\peryr$, this is not long enough to ensure
  a good detection probability. On the other hand, due to the
  relatively high mass of a SBH, such an event can be seen to
  distances as large as a few hundreds of Mpc and will probably dominate
  extra-galactic detection rates.}

  \label{fig:BHCapt}
\end{figure}

The MC simulations yield capture rates for a MW-like galaxy of order
$2\times 10^{-6}$ to $10^{-5}$\,{\peryr} for MSSs, $4\times 10^{-7}$
to $10^{-6}$\,{\peryr} for white dwarfs (WDs) and around $5\times
10^{-8}$\,{\peryr} for neutron stars\footnote{Natal kicks were not
included, so the NS rate is likely overestimated.}  (NSs) and stellar
black holes (SBHs), with large statistical uncertainties for
these rare species.  

To investigate the detectability of a given capture event, we compute
the GW-driven orbital evolution
\cite{GHK02} and GW emission \cite{PPSLR01} and, for a choice of the
distance to the source, $D$, compare the GW amplitude with {\LISA}
sensitivity \cite{LHH00,BH97}. Figures~\ref{fig:MSCapt} and
\ref{fig:BHCapt} illustrate this for a MSS and a SBH event,
respectively. Applying such computations to all capture events during
some time interval, one determines the expected number of captured
stars around {\SgrA} ($D=8\,$kpc) that are emitting above any given
{\em LISA} signal-to-noise ratio (S$/$N). The most striking results
concern MS stars. The predicted number of sources with
$\mathrm{S}/\mathrm{N}\ge 10$ is of order 3-5 if one uses the
optimistic treatment of tidal interactions, and still of order $0.5-2$
with the pessimistic approach. Only MSSs with masses lower than
$0.1\,\Msun$ are resilient enough to tidal forces to contribute. Their
GW signal is too weak to be detected from extra-galactic nuclei. The
probability for a {\SgrA} WD source with $\mathrm{S}/\mathrm{N}\ge 10$
is $0.1-0.5$ and around $0.01$ for NSs and SBHs.

%%%%%%%%%%%%%%%%%%%%%%%%%%%%%%%%%%%%%%%%%%%%%%%%
%% BACKMATTER
%%%%%%%%%%%%%%%%%%%%%%%%%%%%%%%%%%%%%%%%%%%%%%%%

\begin{theacknowledgments}
  The routines to compute the orbital evolution around a MBH were
  kindly provided by Kostas Glampedakis. This work was initiated at
  Caltech, with a fellowship from the Swiss National Science
  Foundation and complementary support from NASA under grant
  NAG5-10707. The work of the author at the Astronomisches
  Rechen-Institut is funded in the framework of project SFB-439/A5 of
  the German Science Foundation (DFG). Financial support from NASA to
  attend this conference is acknowledged.
\end{theacknowledgments}

%% If the bibliography is produced without BibTeX comment out the

\bibliographystyle{aipproc}   % if natbib is available

\bibliography{aamnem99,biblio}

\end{document}